\documentclass[iop]{emulateapj} \pdfoutput=1
\usepackage{amsmath}
\newcommand{\PencilCode}{{ Pencil Code}}
\newcommand{\NDSPHMHD}{{\sc NDSPMHD}}

\begin{document}

\title{A Well-Posed Kelvin-Helmholtz Instability Test and Comparison}
\shorttitle{A Well-Posed Kelvin-Helmholtz Test}

\author{Colin P.~M\textsuperscript{c}Nally\altaffilmark{1}}
\email{{\tt cmcnally@amnh.org}}
\author{Wladimir Lyra\altaffilmark{2,3}}
\email{{\tt wlyra@jpl.nasa.gov}}
\author{Jean-Claude Passy\altaffilmark{4}}
\email{{\tt jcpassy@uvic.ca}}
\affil{Department of Astrophysics, American Museum of Natural History, New York, NY, USA}

\altaffiltext{1}{Department of Astronomy, Columbia University, New
  York, NY, USA}
\altaffiltext{2}{Jet Propulsion Laboratory, California Institute of Technology, 4800 Oak Grove Drive, Pasadena, CA, 91109, USA.}
\altaffiltext{3}{NASA Carl Sagan Fellow}
\altaffiltext{4}{Department of Physics and Astronomy, University of Victoria, Victoria, BC, Canada}

\begin{abstract} 
Recently, there has been a significant level of discussion of
the correct treatment of Kelvin-Helmholtz instability in the
astrophysical community.  This discussion relies largely on how the KHI test is
posed and analyzed.  We pose a stringent test of the initial growth of the
instability.  The goal is to provide a rigorous methodology for verifying a
code on two dimensional  Kelvin-Helmholtz instability.  We ran the problem in
the \PencilCode{}, Athena, Enzo, \NDSPHMHD{}, and Phurbas.  A strict
comparison, judgment,  or ranking, between codes is beyond the scope of this
work, though this work provides the mathematical framework needed for such a
study.  Nonetheless, how the test is posed circumvents the issues raised by
tests starting from a sharp contact discontinuity yet it still shows the poor
performance of Smoothed Particle Hydrodynamics. We then comment on the
connection between this behavior to the underlying lack of zeroth-order
consistency in Smoothed Particle Hydrodynamics interpolation.  We comment on the tendency of some methods,
particularly those with very low numerical diffusion, to produce secondary
Kelvin-Helmholtz billows on similar tests.  Though the lack of a fixed,
physical diffusive scale in the Euler equations lies at the root of the issue,
we suggest that in some methods an extra diffusion operator should be used to
damp the growth of instabilities arising from grid noise.  This statement
applies particularly to moving-mesh tessellation codes, but also to fixed-grid
Godunov schemes.

\end{abstract}


\section{Introduction}

Kelvin-Helmholtz instability (KHI) is the name given to the primary instability
that occurs when velocity shear is present within a continuous fluid or across
fluid boundaries.  The shear is converted into vorticity that, subject to
secondary instabilities, cascades generating turbulence.  
The KHI is one of the most important hydrodynamical instabilities and plays
a significant role in various parts of astrophysics. It is believed to be
responsible for additional mixing in differentially rotating stellar interiors
\citep{2001MNRAS.320...73B}, and to keep a finite-thickness layer of dust
around the midplane of protoplanetary disks
\citep{1995Icar..114..237D,2006ApJ...643.1219J}. 
It also contributes to convective mixing in any
deep stellar interior at the stiff convective boundaries, for instance in
asymptotic giant stars \citep{2006isna.confE.206H} or novae
\citep{2011Natur.478..490C}. Moreover, KHI can lead to the
destruction of cool gravitationally bound objects moving in a hot ambient medium \citep{1993ApJ...407..588M} 
such as
galaxies in the intracluster medium \citep{1982MNRAS.198.1007N,
2000ApJ...538..559M}, substellar companions engulfed by a giant star
\citep{Passy2012inprep} and comets entering a planetary atmosphere
\citep{1994ApJ...434L..33M}.  
KHI plays a role in the interactions of the  
magnetopause and solar wind
\citep{1982JGR....87.7431M} 
and has been observed in the solar corona
\citep{2011ApJ...734L..11O}.
In order to understand these
phenomena and their implications, it is therefore important to define a
well-posed method to quantify how accurately KHI can be modeled by
different numerical techniques.

Verifying the correct treatment of KHI has attracted
increased interest following the conclusions made by \cite{2007MNRAS.380..963A}
including vigorous discussions of KHI in Lagrangian schemes.  The main
conclusion reached was that Smoothed Particle Hydrodynamics (SPH) fails to
resolve KHI due to a surface tension effect between the SPH particles at the
shear interface.  However, the test was done at a sharp shear and contact
discontinuity.  \cite{2008JCoPh.22710040P} attempted to address the problem
with KHI growth from a sharp contact discontinuity in
\cite{2007MNRAS.380..963A} by adding an artificial thermal conductivity to SPH.
A prescription achieving a similar end by adding a diffusion motivated by a
subgrid turbulence model to SPH was proposed by \citet{2008MNRAS.387..427W}.
In a case where traditional SPH largely fails to reproduce KHI at a sharp interface, 
\citet{2010MNRAS.403.1165C} demonstrated that using a Godunov-SPH formulation
with zeroth and first order consistency that growth of KHI can be obtained.  
Using a Voronoi-mesh based scheme
\citet{2010MNRAS.406.2289H} showed improvement over SPH in a sharp contact
discontinuity KHI test, but compared the compressible solution to the growth
rate for an incompressible flow and did not perform a convergence study.  With
the {\sc AREPO}  Voronoi-mesh Godunov code \citet{2010MNRAS.401..791S} ran a
sharp contact discontinuity  KHI test and pointed out the difference seen in
the secondary instabilities developed when the mesh was given a
motion following the flow (a quasi-Lagrangian motion) or kept fixed.  In \cite{2011arXiv1109.2218S} the same
code is used for a extended discussion, with both a sharp contact discontinuity
KHI test and a smooth transition test, but comparing both of the compressible
results to the growth rate for an incompressible sharp contact continuity
initial test.  \citet{2010MNRAS.405.1513R} pointed out the zeroth-order
inconsistency in SPH, and designed a kernel to minimize these effects,
achieving better qualitative results on a sharp contact discontinuity KHI test.
Zeroth-order inconsistency is the inability of SPH interpolation to reproduce a
constant function at any finite resolution
\citep{LiuJunZhang1995,Dilts1999,LiuLiuLam2003,Fries2004,Quinlan2006}.
A quantitative analysis of the growth of KHI from a sharp contact discontinuity
was performed by \citet{2010MNRAS.407.1933J} with SPH and grid-based Godunov
codes.  With a focus on SPH, \citet{2010MNRAS.408...71V} argued that a sharp
contact discontinuity KHI test is not ideal, and propose an alternative SPH
smoothing kernel which yields improved results.  \cite{2011IAUS..270..429H}
compared qualitatively a grid based method and SPH using both a cubic and
quintic kernel on a sharp contact discontinuity KHI test, finding that the
choice of a quintic kernel improved the SPH results significantly.  One of the
only well posed convergence tests for KHI was done in
\cite{2010MNRAS.401.2463R}, but that was in a study of Galilean invariance
restricted to fixed-mesh schemes.  The test by \cite{2010ARA&A..48..391S} is a
well posed problem influenced by \cite{2010MNRAS.401.2463R}, but the evaluation
of the SPH result was done by comparison to an
analytic solution for a sharp transition initial condition and incompressible
flow in an infinite domain, not for the problem posed with a softened
transition in compressible flow in a finite periodic domain.  

The commonly used solution for the KHI growth rate in numerical tests is for a sharp
transition at the shear interface \citep[][sec.~101]{chandrasekhar}.  
However, for numerical approximations the interface should
be smoothed to yield an initial value problem with finite spatial derivatives,
as argued by \cite{2010MNRAS.401.2463R}.  For a sharp interface,
the initial approximation of the derivatives across the interface does not
converge with resolution. To obtain convergence a smooth interface must be
used.  We pose the problem in such a way that the analytic result for the
incompressible limit is known for an infinite domain, as this type of analytic
result is usually used to compare numerical results.  However, a difficulty
with Kelvin-Helmholtz problems is that the unstable modes are global, so
solutions in a finite periodic domain, as commonly tested, are different from
than the solution in an infinite domain.  To circumvent this difficulty, we
perform an exhaustive convergence study to establish a fully compressible
nonlinear solution with a very small and rigorously
derived uncertainty.

In the following discussion, we will refer to different types of
discretizations used for numerical solutions to the chosen governing equations
of hydrodynamics or magnetohydrodynamics.  To clarify, most fixed grid (or
structured mesh) codes use a square Eulerian grid as a basis for either a
point-value (values of the fields at grid points) or volume-average (average
volume of the field in a grid cell) discretization.  The distinction here is
that a finite-volume scheme can be arranged to solve the integral form of the
governing equations, and the point-value discretization can only solve the
differential form of the equations.  Unstructured mesh discretizations do not
pose the restriction of the discretization mesh being a regular grid, however
the  nodes are logically connected by edges, and the mesh cells form a
tessellation on the computational volume.  Moving-mesh Voronoi tessellation
discretizations have begun to appear in Astrophysical applications \citep{2010MNRAS.401..791S,2011arXiv1104.3562D}.
These are
a case of an unstructured mesh, where the mesh is defined by the Voronoi
tessellation of a set of mesh generating points. The mesh cells may be used to
define a finite-volume discretization.  When the mesh generating points are
allowed to move in time to make a moving-mesh discretization, the Voronoi mesh
will be recalculated on the new point distribution at every step yielding a new
set of cells and new mesh edges connecting them.  The mesh movement can be
arbitrary, but a quasi-Lagrangian mesh movement is a particularly good choice
as this minimizes numerical errors associated with advection across the grid.
It is also possible to define meshless discretizations that represent the
fields on a set of points or particles without specifying a set of mesh edges
to connect these points.  Meshless discretizations then do not form a strict
tessellation of the computational volume.  These discretizations are commonly
used to define Lagrangian methods, in the sense that the points or particles
are comoving with the fluid.  If finite-mass particles are used, then the
particles form a partition of the total computational mass, and the form of
discretization used in SPH is obtained and the
integral form of the governing equations can be solved.  However, if the
meshless points carry only field values at those points, then a point-value
type discretization is obtained and the differential form of the governing
equations must be solved.

In Section~\ref{sec_setup} we give the problem setup used and in
Section~\ref{sec_analysis} we discuss the methods used to extract measured
quantities from the results.  The codes used in this paper are listed in
Section~\ref{sec_codes}.  The detailed convergence study used to generate the
reference compressible solution is presented in Section~\ref{sec_reference}. 
The results and comparison
from several codes with different underlying algorithms and discretizations are
presented in Section~\ref{sec_results}.  We discuss the various results and
implications in Section~\ref{sec_discussion}.  Extended discussion of the SPH
results and analysis of extra experiments is in Section~\ref{sec_sph}.
In Section~\ref{sec_secondary} we discuss secondary instabilities
arising from the problem setup in this work, and the difficulty of determining
if they are produced in a physically meaningful manner.  Our conclusions are
summarized in Section~\ref{sec_conclusions}.

\section{Setup} \label{sec_setup}

\begin{figure*}
\begin{center}
\plotone{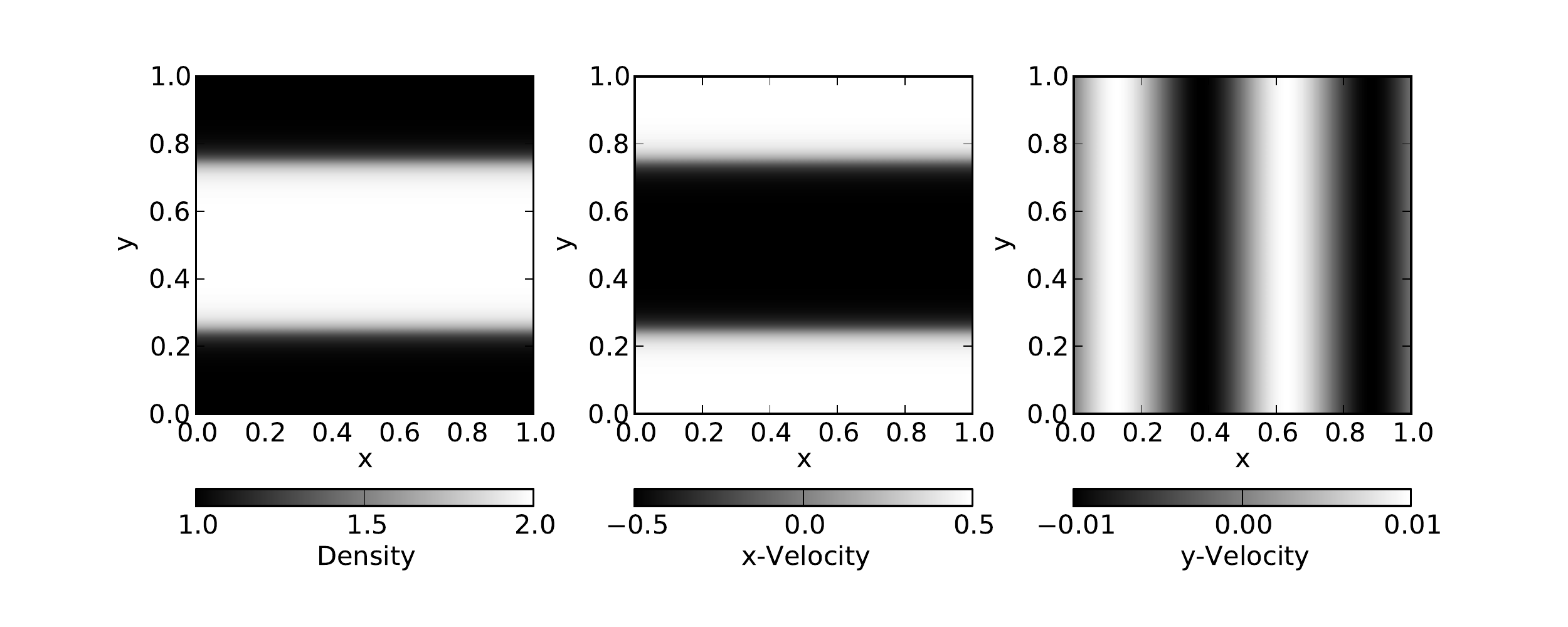}
\caption{Density and velocity initial conditions used for density and velocity for the KHI test in this work.} 
\label{figic}
\end{center}
\end{figure*}

Our motivation in choosing the initial condition is that the initial conditions
are smooth, reflect as closely as possible a configuration that can be treated
analytically, and can be represented easily in a wide variety of codes.  In all
codes, we will solve the inviscid compressible Euler equations.  The setup we
use is chosen to be a periodic version of that used in the analysis of
Kelvin-Helmholtz instability in \cite{2010PhPl...17d2103W}: the domain is 1
unit by 1 unit in the $x$ and $y$ directions if two-dimensional, and an
arbitrary thickness in the $z$ direction if needed for a three dimensional
code.  Runs with resolutions of $128\times 128$, $256\times256$, and
$512\times512$ cells, or equivalent were used in the comparison.  All
boundaries are periodic.  The initial condition is smooth and periodic, as
illustrated in Figure~\ref{figic}.  The density is given by:
\begin{equation}
\rho = \left\{ \begin{array}{ll}
\rho_1 - \rho_m e^\frac{y-1/4}{L} &\mbox{ if } 1/4>y\geq 0\\
 \rho_2 +\rho_m e^\frac{-y+1/4}{L} &\mbox{ if } 1/2>y\geq 1/4\\
 \rho_2 +\rho_m e^\frac{-(3/4-y)}{L} &\mbox{ if } 3/4>y\geq 1/2\\
 \rho_1 -\rho_m e^\frac{-(y-3/4)}{L} &\mbox{ if } 1>y\geq 3/4
\end{array}
\right.
\end{equation}
where
\begin{equation}
\rho_m = (\rho_1 - \rho_2)/2
\end{equation}
with $\rho_1 = 1.0$, $\rho_2 = 2.0$, and the smoothing parameter $L=0.025$.
The $x$-direction velocity is given by:
\begin{equation}
V_x = \left\{ \begin{array}{ll}
U_1 - U_m e^\frac{y-1/4}{L} &\mbox{ if } 1/4>y\geq 0\\
U_2 + U_m e^\frac{-y+1/4}{L} &\mbox{ if } 1/2>y\geq 1/4\\
U_2 + U_m e^\frac{-(3/4-y)}{L} &\mbox{ if } 3/4>y\geq 1/2\\
U_1 - U_m e^\frac{-(y-3/4)}{L} &\mbox{ if } 1>y\geq 3/4\\
\end{array}
\right.
\end{equation}
where
\begin{equation}
U_m = (U_1 - U_2)/2
\end{equation}
with $U_1 = 0.5$, $U_2 = -0.5$, and $L$ as in the density so that
 the smooth transition in density and velocity occurs over the same interval.
The background shear is perturbed by adding some velocity in the $y$-direction with the form
\begin{align}
V_y = 0.01\sin(4\pi x).
\end{align}
An ideal gas equation of state  with $\gamma=5/3$ is used. The internal energy
is set such that pressure is initially uniform with value $2.5$. The problem is
run till at least time $t=1.5$.  Analysis is done on snapshots spaced at a
minimum of $\Delta t=0.02$.  However, in most cases the snapshots will not be
spaced exactly as  codes often do output or analysis on an approximate
interval, e.g.\ at the first time step after the specified snapshot or analysis
time.  The test can be run in two dimensions in a structured grid code, but for
unstructured meshes or mesh free methods two dimensional and three dimensional
simulations may yield slightly different results depending on how the resolution
elements are arranged in the initial condition.  For unstructured mesh methods
and meshless methods the results will differ for a disordered node distribution
and a regularly gridded one.

\section{Analysis} \label{sec_analysis}
To quantitatively describe the growth of the Kelvin-Helmholtz instability, we
use two measurements, the amplitude of the  $y$-velocity mode of the
instability, and the maximum $y$-direction kinetic energy density.  These two
quantities are a useful pair, as the mode amplitude is a smoothed quantity and
the maximum $y$-direction kinetic energy density is very sensitive to noise in
the computed velocity field.

As a loose guide, the analysis of \cite{2010PhPl...17d2103W} treats a
non-periodic incompressible version of the problem studied in this paper. Their
linear perturbation theory yields growth rates for the two quantities studied in
this work, in the infinite domain and incompressible flow limit. However, as we
run our test in periodic boundaries with a compressible flow we must go further
than their analysis.

The maximum $y$-direction kinetic energy is the simplest of the
two quantities to compute.  This quantity is  the maximum value of  $1/2\rho
V_y^2$ computed for all resolution elements (cells, points, or particles) in
the computation volume at each time.  In the non-periodic, incompressible
limit, the growth of this quantity should be $\propto \exp(2\times4.384\times t)$
\citep[][Equation 18]{2010PhPl...17d2103W}.  In practice, the growth will start from
a finite perturbation,  will reflect erroneous velocities occurring both at the
interface due to unbalanced pressures at the cell scale, and any velocity and
density noise in the bulk flow.  It is also important that the test posed here,
and those commonly used in other works, are actually posed in a periodic domain
with a compressible flow.  To obtain a basis for comparison we use a numerical
reference solution to the problem as posed and establish the uncertainty on
this reference solution in a rigorous manner in Section~\ref{sec_reference}.

To extract the amplitude of the $y$-velocity mode of the instability a more
involved calculation is required.  We wish to define the measurement in a
manner that can be made consistent across different types of discretizations.
A simple Fourier transform defined on a grid would be entirely appropriate for
point-based finite difference schemes or pseudospectral schemes, but is
somewhat less well motivated for finite-volume schemes, and inappropriate for
meshless or unstructured mesh schemes.  To state the analysis in a manner that
is straightforward to describe for all codes, which treats all results in the
same manner we use a discrete convolution.  For the case of a uniform grid this
amplitude $M$ is given by: 
\begin{align}
s_i &=\left\{\begin{array}{l l} V_y\sin(4\pi x)\exp(-4\pi|y-0.25|) & \text{if } y<0.5\\
 V_y\sin(4\pi x)\exp(-4\pi|(1-y)-0.25|) & \text{if }y \geq 0.5\label{eqmodebegin}\\
\end{array}\right. \\
c_i &=\left\{\begin{array}{l l} V_y\cos(4\pi x)\exp(-4\pi|y-0.25|) & \text{if } y<0.5\\
 V_y\cos(4\pi x)\exp(-4\pi|(1-y)-0.25|) & \text{if }y\geq 0.5\\
\end{array}\right. \\
d_i &=\left\{\begin{array}{l l} \exp(-4\pi|y-0.25|) & \text{if } y<0.5\\
 \exp(-4\pi|(1-y)-0.25|) & \text{if }y\geq 0.5\\
\end{array}\right. \\
M &= 2\sqrt{\left(\frac{\sum_{i=1}^N{s_i}}{\sum_{i=1}^N{d_i}}\right)^2+\left(\frac{\sum_{i=1}^N{c_i}}{\sum_{i=1}^N{d_i}}\right)^2} \label{eqmodeend}
\end{align}
 where $i$ ranges over all grid points or cell centers ($N$ total grid points
or cell centers) and the positions $(x,y)$ are grid points or cell centers.  
This expression can be used in two or three dimensions.
The mode amplitude $M$ is calculated at each time snapshot.

For SPH simulations, each particle needs a different weighting in the sums as
the particle density varies.  In the case of variable smoothing length SPH
where the smoothing length is set to encompass a fixed number of neighbors, we
can use the smoothing length $h_i$ for particle $i$ to do this weighting. In
the following formulas $p$ is the number of dimensions the simulation is run
in.
\begin{align}
s_i &=\left\{\begin{array}{l l} V_y h_i^p \sin(4\pi x)\exp(-4\pi|y-0.25|) & \text{if } y<0.5\\
 V_y h_i^p \sin(4\pi x)\exp(-4\pi|(1-y)-0.25|) & \text{if }y\geq 0.5\\
\end{array}\right. \\
c_i &=\left\{\begin{array}{l l} V_y h_i^p \cos(4\pi x)\exp(-4\pi|y-0.25|) & \text{if } y<0.5\\
 V_y h_i^p \cos(4\pi x)\exp(-4\pi|(1-y)-0.25|) & \text{if }y\geq 0.5\\
\end{array}\right. \\
d_i &=\left\{\begin{array}{l l} h_i^p  \exp(-4\pi|y-0.25|) & \text{if } y<0.5\\
 h_i^p  \exp(-4\pi|(1-y)-0.25|) & \text{if }y\geq 0.5\\
\end{array}\right. \\
M &= 2\sqrt{\left(\frac{\sum_{i=1}^N{s_i}}{\sum_{i=1}^N{d_i}}\right)^2+\left(\frac{\sum_{i=1}^N{c_i}}{\sum_{i=1}^N{d_i}}\right)^2}
\end{align}
The quantities $s_i$ and $c_i$ are defined for each particle $i=1..N$, from the
position $(x,y)$ and the $y$-velocity $V_y$ of that particle.  An advantage of
the definition used here is that we can directly analyze the SPH particle
values as simulated without introducing an additional interpolation to a fixed
grid. This feature follows over to a unstructured mesh or meshless code.

For an unstructured mesh code, or a meshless code that defines quadrature
volumes for the points the appropriate general form would be:
\begin{align}
s_i &=\left\{\begin{array}{l l} V_y w_i \sin(4\pi x)\exp(-4\pi|y-0.25|) & \text{if } y<0.5\\
 V_y w_i \sin(4\pi x)\exp(-4\pi|(1-y)-0.25|) & \text{if }y\geq 0.5\\
\end{array}\right. \\
c_i &=\left\{\begin{array}{l l} V_y w_i \cos(4\pi x)\exp(-4\pi|y-0.25|) & \text{if } y<0.5\\
 V_y w_i \cos(4\pi x)\exp(-4\pi|(1-y)-0.25|) & \text{if }y\geq 0.5\\
\end{array}\right. \\
d_i &=\left\{\begin{array}{l l} w_i  \exp(-4\pi|y-0.25|) & \text{if } y<0.5\\
 w_i  \exp(-4\pi|(1-y)-0.25|) & \text{if }y\geq 0.5\\
\end{array}\right. \\
M &= 2\sqrt{\left(\frac{\sum_{i=1}^N{s_i}}{\sum_{i=1}^N{d_i}}\right)^2+\left(\frac{\sum_{i=1}^N{c_i}}{\sum_{i=1}^N{d_i}}\right)^2}
\end{align}
where $w_i$ is the area or volume of cell or the quadrature volume for point
$i$ and the positions are the cell centers or point positions.

For an infinite domain with incompressible flow, the growth of the velocity
mode $M$ should be  $\propto \exp(4.384\times t)$
\citep[][Eq.~18]{2010PhPl...17d2103W}.  The growth rate for a Kelvin-Helmholtz
instability with these two conditions has been used before as a comparison for
results obtained in periodic domains with a compressible flow, but the two
problems are formally different, and depending on the parameters the growth
rates may vary.  Again, to circumvent this difficulty, we compare results to
the numerical solution of the test problem specified, and establish the
uncertainty on this reference solution in a rigorous manner in
Section~\ref{sec_reference}.

\section{Codes} \label{sec_codes}

\begin{table*}
\begin{center}
\caption{Simulation Prefixes and Codes}
\label{prefixtable}
\begin{tabular}{lll}
\hline
\hline
Prefix & Code & Variation \\
\hline
Pe & Pencil & 6th order space, 3rd order time accuracy, 6th-order hyperviscosity \\
Ep & Enzo & 3rd order reconstruction, directionally split, two-shock Riemann solver\\
At & Athena & 3rd order reconstruction, unsplit integrator, HLLC Riemann solver\\
Ne & \NDSPHMHD{} & 2D cubic kernel\\
Nc & \NDSPHMHD{} & 2D cubic kernel, no artificial conductivity\\
No & \NDSPHMHD{} & 2D quintic kernel\\
Ph & Phurbas  & 3D, $\lambda$ = cell length in 2D codes\\
\hline
\end{tabular}
\end{center}
\end{table*}

In this paper, we compare the results from six codes to the reference solution,
which itself is produced with the {\sc Pencil code}.

The {\sc Pencil code}{\footnote{See {\tt
http://www.nordita.org/software/pencil-code}}} is a fixed Eulerian mesh,
non-conservative, finite-difference, MHD code that uses sixth order centered
spatial derivatives and a third order Runge-Kutta time-stepping scheme, being
primarily designed for weakly compressible turbulent hydromagnetic flows.  For
the problem in question, in order to keep the Reynolds number low at the grid
scale while keeping the integral and intermediate scales nearly inviscid,
explicit sixth-order hyperdiffusion and hyperviscosity are added to the mass
and momentum equations as specified in \cite{2008A&A...479..883L}
{\footnote{For reproducibility purposes, we quote the hyperviscosity type,
value, and the svn revision number of the {\sc Pencil Code} version used. The
version was r17470 or thereabouts, the diffusion was set to {\tt
idiff=``hyper3-mesh''}, with hyperdiffusion coefficient {\tt
diffrho\_hyper3\_mesh}=20. The viscosity type was set to {\tt
ivisc=``hyper3-mesh''} with hyperviscosity coefficient {\tt
nu\_hyper3\_mesh}=20. The coefficients are inversely proportional to the grid
Reynolds number.}}.

The other codes, Enzo, Athena, \NDSPHMHD{} and Phurbas are introduced below.

Enzo is a three-dimensional, Eulerian adaptive mesh refinement hybrid
(hydrodynamics + N-body) grid-based code
\citep{1995AAS...187.9504B,2004astro.ph..3044O}\footnote{\tt
http://enzo-project.org/}.  For this problem the Euler equations are solved
using a third-order piecewise parabolic method (PPM) with the two-shock
approximate Riemann solver.  Time-stepping is constrained by a Courant
condition for the gas with a Courant factor C=0.4.  The run-time PPM diffusion,
flattening, and steepening parameters were set to zero.  Enzo version 1.5 was
used.

Athena is a three dimensional Eulerian grid code that (among other algorithms)
implements a higher order Godunov method for hydrodynamics
\citep{2008ApJS..178..137S}.  Specifically, we have used the third-order cell
reconstructions with the HLLC approximate Riemann solver and the unsplit
corner-transport-upwind (CTU) second order time integration algorithm.
Otherwise the options used were as specified in the two-dimensional test
problem supplied with the code, with a Courant number $C=0.8$.  We used Athena
version 4.1 obtained from the project website\footnote{{\tt
https://trac.princeton.edu/Athena/}}.

\NDSPHMHD{} is a one, two, and three dimensional reference implementation of
SPH and a platform for experimentation
\citep{2012JCoPh.231..759P}.  We obtained \NDSPHMHD{} version 1.0.1 from the
author's website\footnote{{\tt
http://users.monash.edu.au/\textasciitilde{}dprice/ndspmhd/index.html}}.
\NDSPHMHD{} was run on this problem in two dimensions, using both the cubic and
quintic kernel options.  The cubic kernel is the conventional choice for SPH,
whereas the quintic kernel delivers higher accuracy at the cost of
computational expense.  \cite{2012JCoPh.231..759P} describes the \NDSPHMHD{}
implementation of SPH as converging as higher order kernels are used.  That is,
the result on the test problem shown here should converge with the combination
of using more particles and using a higher order kernel.  \NDSPHMHD{} also
supports the artificial thermal conductivity described in
\cite{2008JCoPh.22710040P}.  The results of SPH simulations may depend strongly
on the initial particle distribution used.  
To generate the initial condition, we relaxed a set of equal mass particles to
an approximate equilibrium with an artificially imposed pressure field which
produced the required density profile.  The particles settle into a roughly
hexagonal grid, although with dislocations required to produce the spatially
varying density.
The number of particles used at each resolution matched the number of cells or
points used for the grid code ($128^2$, $256^2$, $512^2$).  Otherwise,
the code was run with the default parameters  used in Test 6 of the \NDSPHMHD{}
examples package.

Phurbas is a meshless, adaptive, Lagrangian code for magnetohydrodynamics
\citep{2011arXiv1110.0835M,2011arXiv1110.0836M}.  Phurbas uses third order
least square fits to derive spatial derivatives, and a second order scheme for
time integration.  Stabilization is achieved through an artificial bulk
viscosity.  It is run here in three dimensions, using volumes with height
$1/64$, $1/128$, and $1/256$ in thickness in the $z$-direction.  Phurbas does
not use a grid, so instead we use spatially constant resolution and set the
resolution parameter $\lambda$ to the cell size used in the grid codes.  To
produce the initial particle distribution, we first used a tiling procedure as
in \cite{2011arXiv1110.0836M} and then further relaxed the distribution to one
that would arise naturally in a shearing flow by running the problem to $t=1.5$
and restarting the test with the initial condition defined on resulting
particle distribution.  Because the disordered particle distribution is
inherently three dimensional, the results at a given resolution cannot be
strictly compared to the two-dimensional runs performed in other codes here.  

\section{A Reference Solution of Known Quality} \label{sec_reference}

\begin{figure}
\begin{center}
\includegraphics[width=5cm]{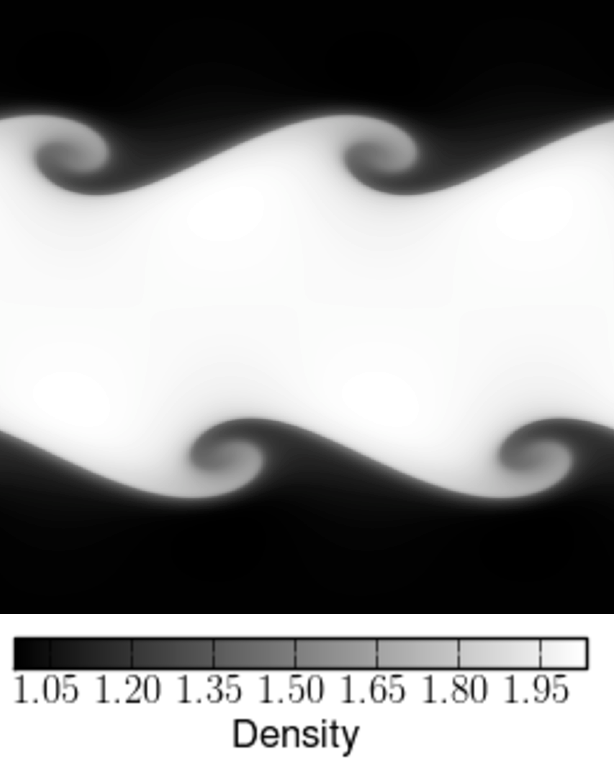}
\caption{Density in the highest resolution Pencil Code simulation used for the reference solution, grid size $4096^2$ output at time $t=1.5$.} 
\label{figpencilkhrho4096}
\end{center}
\end{figure}

\begin{figure}
\begin{center}
\plotone{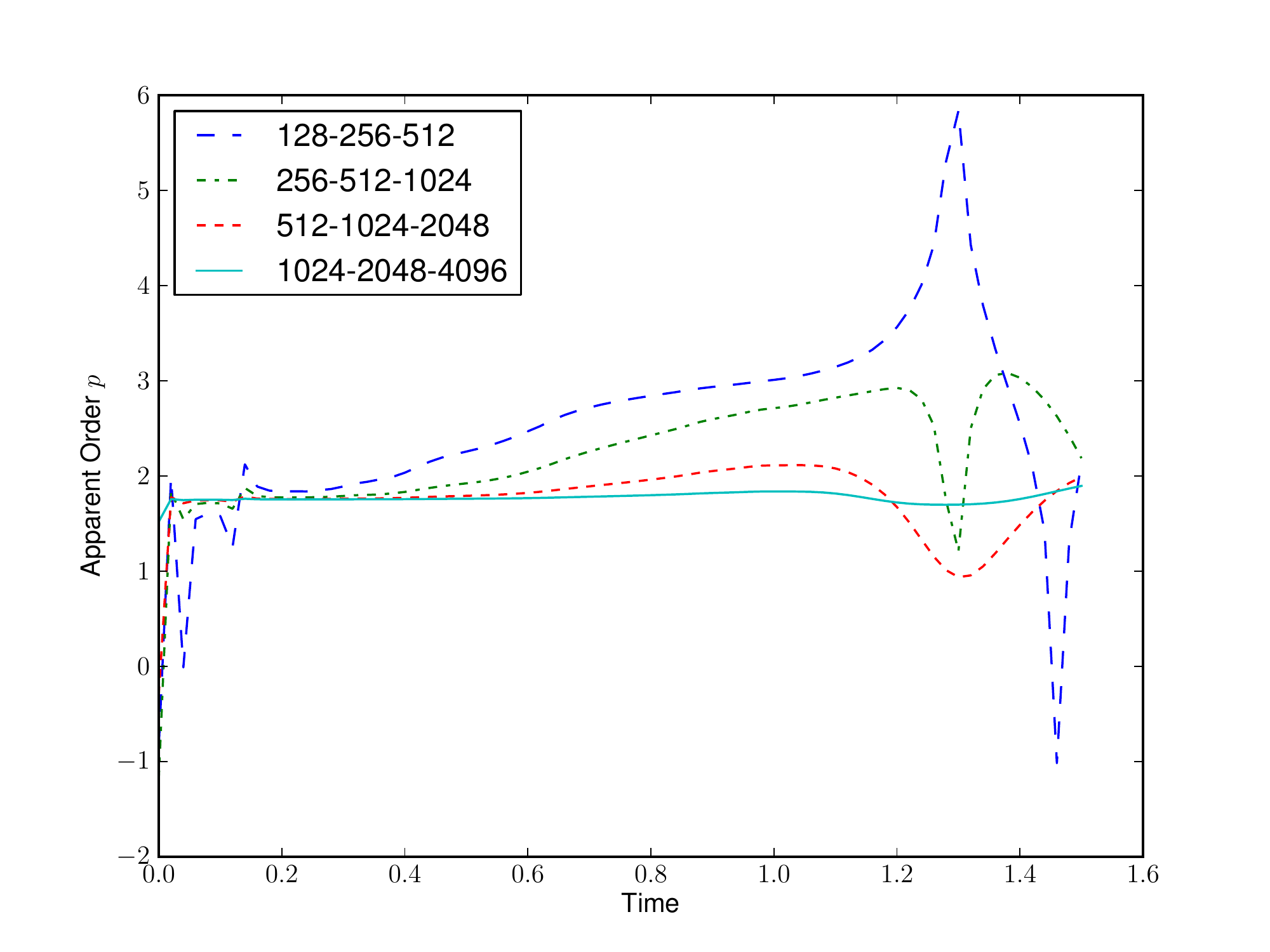}
\caption{Pencil Code apparent method order/rate of convergence measured over each set of three resolutions as denoted in the legend. 
The apparent order converges towards 1.75.}
\label{empiricalorderplot}
\end{center}
\end{figure}

\begin{figure}
\begin{center}
\plotone{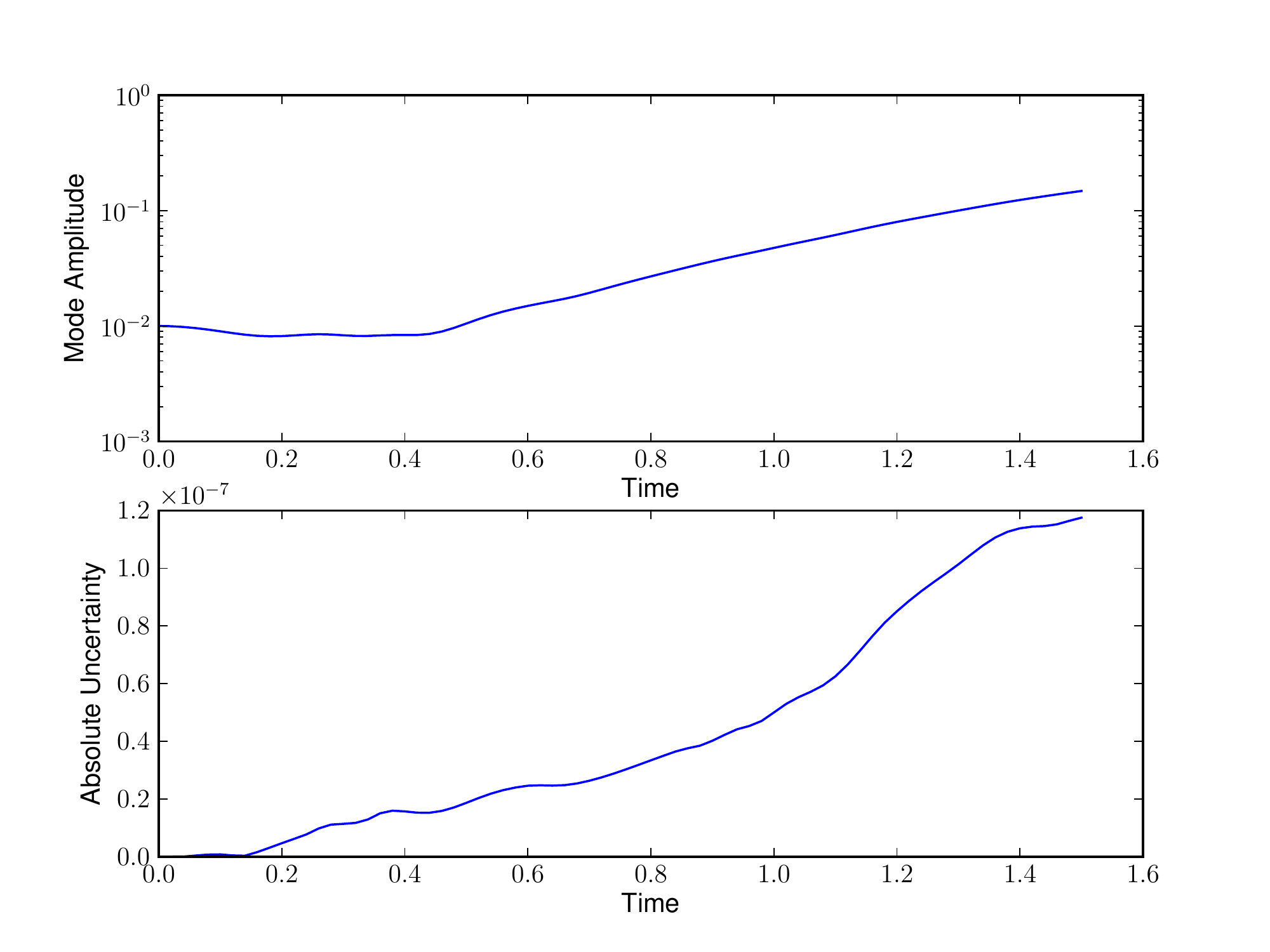}
\caption{Pencil Code  result at $4096\times 4096$ grid points, and uncertainty derived by Richardson Extrapolation based Grid Convergence Index.}
\label{pencilrefsolplot}
\end{center}
\end{figure}

\begin{figure*}
\begin{center}
\plotone{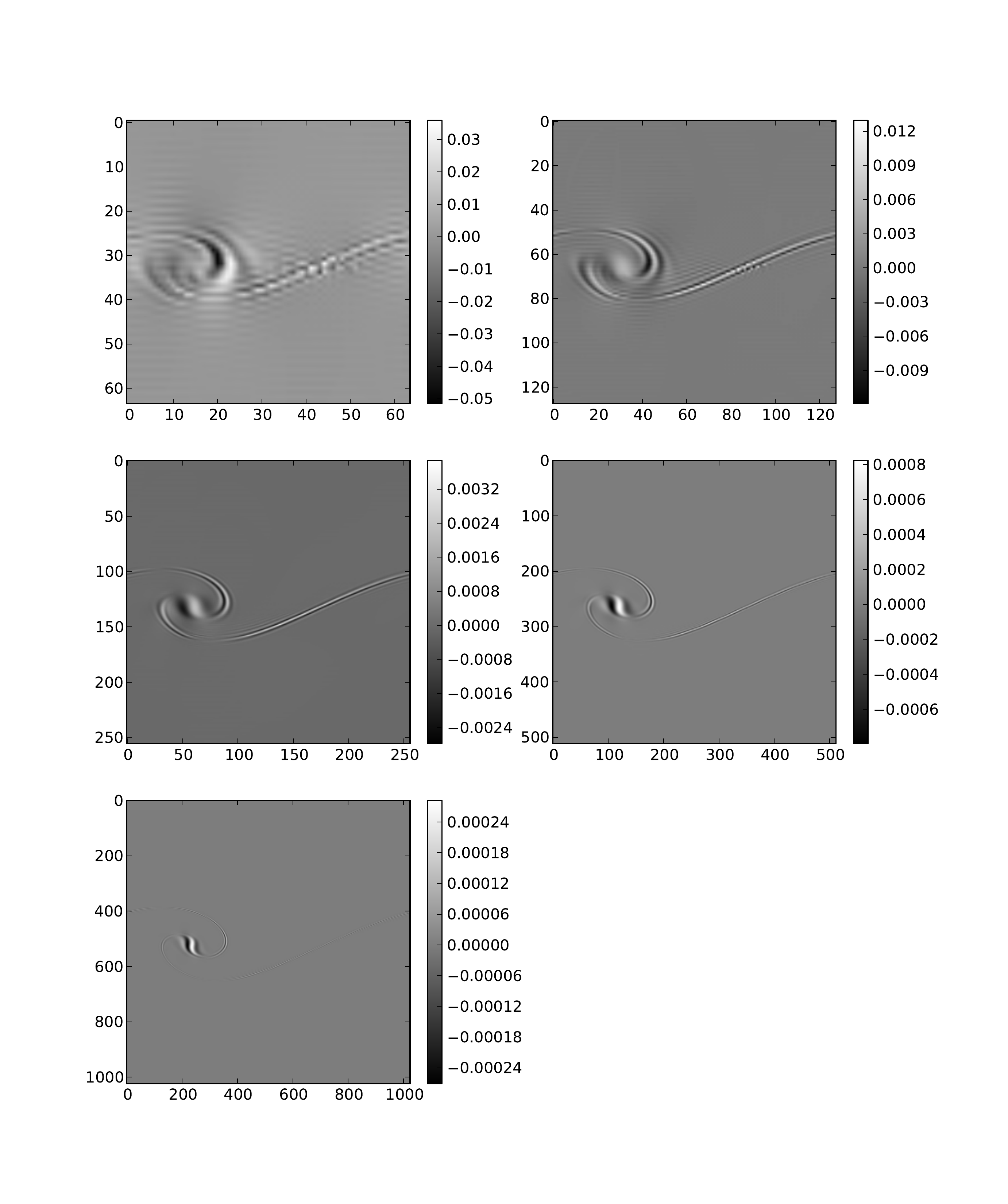} \caption{Differences in $y$-velocity between
sucessively finer resolutions in one quadrant of the convergence study
performed with the Pencil Code at time $t=1.5$. 
Color bars show range of $y$-velocity differences, and 
axes are in units of grid points in the lower resolution for each plot.
Differences shown are:
{\em Upper Row:} $128^2$-$256^2$ and $256^2$-$512^2$
{\em Middle Row:} $512^2$-$1024^2$ and $1024^2$-$2048^2$
{\em Lower Row:} $2048^2$-$4096^2$}
\label{fig_refvydiff}
\end{center}
\end{figure*}

\begin{figure*}
\begin{center}
\plotone{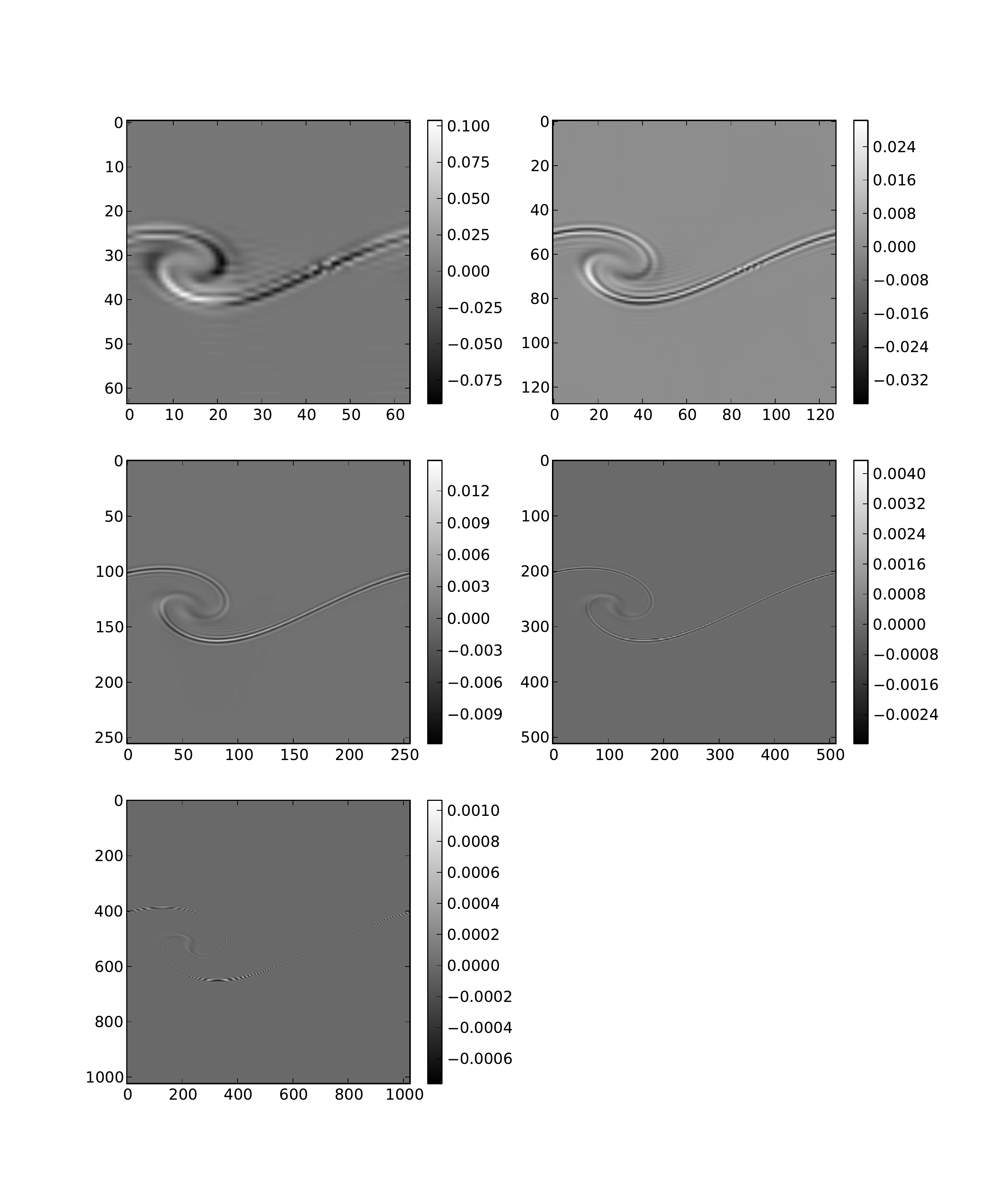}
\caption{
Differences in density between successively finer resolutions in one quadrant of
the convergence study performed with the Pencil Code at time $t=1.5$.  Color
bars show range of density differences, and axes are in units of grid points in
the lower resolution for each plot.  Differences shown are:
{\em Upper Row:} $128^2$-$256^2$ and $256^2$-$512^2$
{\em Middle Row:} $512^2$-$1024^2$ and $1024^2$-$2048^2$
{\em Lower Row:} $2048^2$-$4096^2$}

\label{fig_refrhodiff}
\end{center}
\end{figure*}

To produce a solution to the full nonlinear, periodic, compressible case as run
in this work, we performed an extensive convergence study with the \PencilCode{}.
This convergence study allows us to establish not only a very high quality
reference solution, but also a notion of the uncertainty in this reference
solution.  The importance of the unusual step of establishing the uncertainty
of the reference result is that we can then assert with confidence that the
differences seen between other lower quality results and this reference result
are overwhelmingly due to errors in the lower quality solutions.  In the
results in Section~\ref{sec_results} the Pencil Code is shown to be well suited to
the smooth, subsonic problem posed here.  We use grids of $128\times 128$, $256\times 256$,
$512\times 512$, $1024\times 1024$, $2048\times 2048$, and $4096\times 4096$
points, specified so that every second grid coordinate overlaps on successive
refinements, and with the time stepping scheme in the Pencil Code modified to
provide outputs at exact $\Delta t = 0.02$ time unit intervals.  This set of outputs
enables a resolution study at each output time for the convergence of the mode
amplitude.  Establishing the empirical rate of convergence of the mode
amplitude $M$ allows a Richardson extrapolation based estimate of the
uncertainty in the most resolved measurement.  Hence, we are able to make
comparisons of the results from other codes to the highest resolution Pencil
Code result while knowing in a rigorous manner that the errors in this
reference result are negligible.  

First, we can calculate the empirical rate of
convergence $p$  of the mode amplitudes defined by
Equations~\ref{eqmodebegin}-\ref{eqmodeend} for a set of three results with a
refinement ratio of 2 between each resolution as \begin{equation}
p = \ln\left(\frac{f_3-f_2}{f_2-f_1}\right)/\ln(2)
\label{eqorder}
\end{equation}
where $f_3$ is the value of the mode amplitude on the coarsest grid and $f_2$,
$f_1$ are the values on the medium and finest grid respectively \cite[][Equation~5.10.6.1]{roache}.  
This rate of convergence tells us how fast the series of
values from each resolution is converging towards the correct result.  Once we
have identified the convergence rate of the series of results, we can apply a
generalized form of Richardson extrapolation to estimate the converged result,
and hence derive an indication of the uncertainty in our highest resolution
result.  This indication of uncertainty is the Grid Convergence Index
\citep[GCI,][]{roache}, a uniform method of reporting the uncertainty on such a
convergence study given as \begin{equation}
GCI = F_s\frac{(f_2-f_1)/f_1}{1-r^p}
\label{eqgci}
\end{equation}
from \citet[][Equation~5.6.1]{roache}. The value of the safety factor $F_s$ we use is
$1.25$.
This value is that suggested by \citet[][Section~5.9]{roache} as being appropriate 
when the rate of 
convergence is explicitly determined with a
convergence study as in this work.

A density plot at time $t=1.5$ from the highest resolution ($4096^2$)
calculation is shown in Figure~\ref{figpencilkhrho4096}.  The results of
evaluating Equation \ref{eqorder} for each set of three resolutions is shown in
Figure \ref{empiricalorderplot}. This figure shows that the convergence rate
settles at approximately $1.75$ for most of the time interval $t=0-1.5$ when
the highest resolution results are considered.  Using the observed rate of
convergence at each time, we can assign the uncertainty on the result with
Equation~\ref{eqgci}, which is shown in Figure~\ref{pencilrefsolplot}.  The
high resolution results used are necessary to establish a well behaved
convergence in Figure~\ref{empiricalorderplot}, which means that the
uncertainty is only well known when the uncertainty itself is very small.  To
demonstrate more explicitly the convergence behavior, and the magnitude of the
changes between successive resolutions we have plotted the differences in the
$y$-velocity values between successive resolutions in one quadrant of the
domain in Figure~\ref{fig_refvydiff}.  The greatest changes between successive
resolutions are  localized to the density change interface, and show not
suggestion of the presence of secondary instabilities.  A similar plot for the
density is shown in Figure~\ref{fig_refrhodiff}, again showing no suggestion of
secondary instabilities.

\section{Results} \label{sec_results}

The simulations are identified by a two letter prefix as outlined in Table
\ref{prefixtable} and the resolution ($128^2,$ $256^2$, $512^2$).  Results for the
$y$-velocity unstable mode amplitude itself, and the growth of that quantity
for all codes are plotted in  Figure~\ref{fig7.pdf}.
Figure~\ref{allenergyplot} gives the results for all codes for the minimal
$y$-direction kinetic energy. 
The following two subsections discuss these two measured quantities.

\subsection{$y$-Velocity Unstable Mode Amplitude}
\begin{figure*}
\begin{center}
\includegraphics[width=16cm]{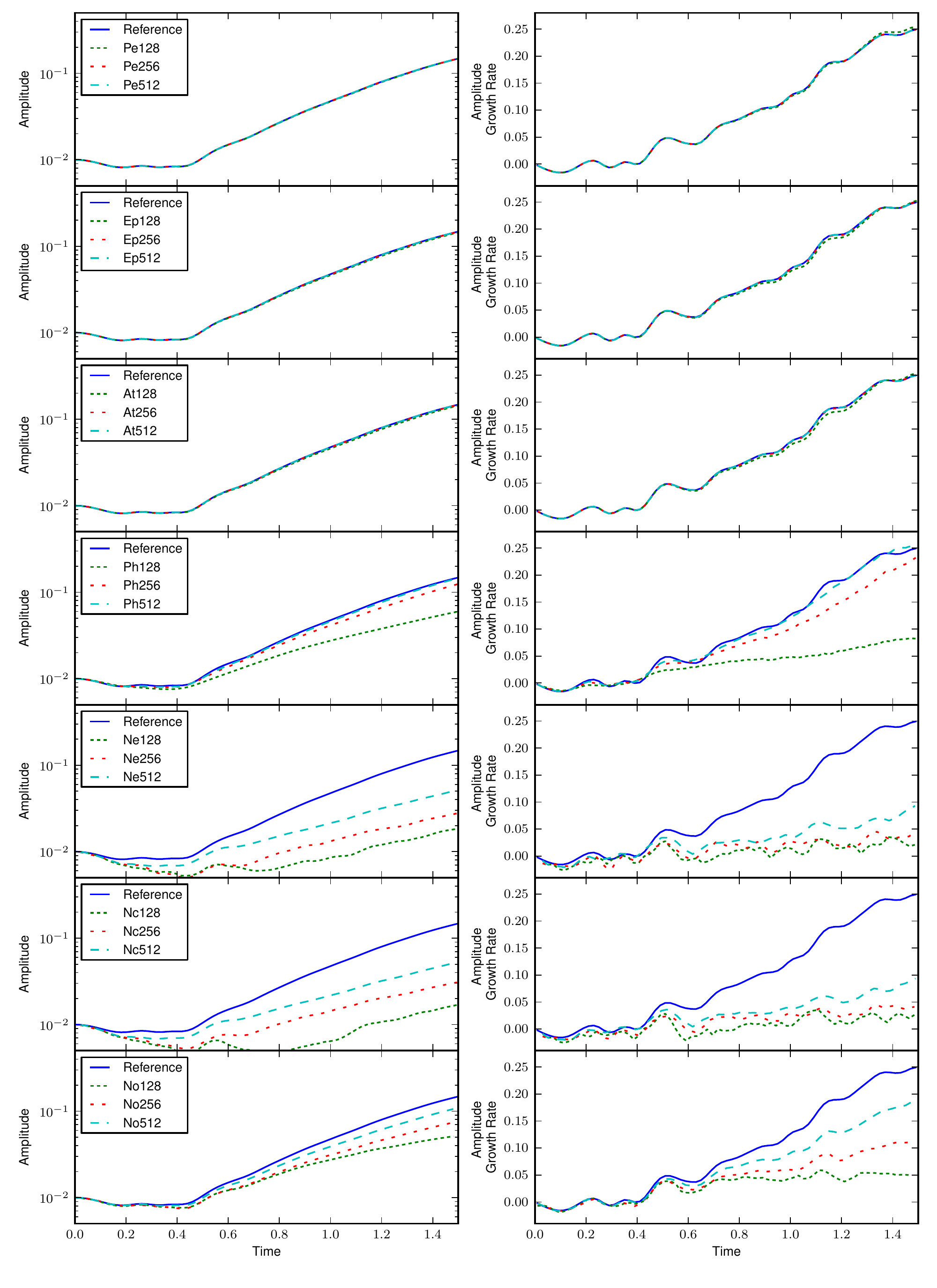}
\caption{Mode growth in all codes.}
\label{fig7.pdf}
\end{center}
\end{figure*}

In interpreting the $y$-velocity unstable mode amplitude
(Figure~\ref{fig7.pdf}) it is important to note that a relative comparison
of the solution quality of the codes cannot be made exactly.  For the
unstructured mesh and meshless methods, the code performance in two dimensions
and three dimensions is expected to differ notably as the possible arrangements
of cells and particles differs.  A strict comparison between Phurbas and the
other codes cannot be drawn as Phurbas was run in three dimensions  not two.
For Eulerian grid codes, the problem is grid-aligned, and performance will
differ as it is rotated against the grid.  With these caveats, we proceed to
comment on the results obtained.

The results for the growth of the $y$-velocity
unstable mode amplitude in the Pencil Code, Enzo, and Athena are very similar
at the level of this comparison.  Here, the  main difference is a variation
between the codes of the growth rate at the lowest resolution.  Reassuringly,
the $128^2$ resolution mode amplitude growth curves from the two piecewise
parabolic method variations used in Enzo and Athena resemble each other more
than they do the result from the Pencil Code.  These results demonstrate that
the Pencil Code reference result is reasonable.

For Phurbas  unstable mode amplitudes  converge with increasing
resolution from below the reference value, but at the $512^2$ resolution the
growth rate at late times exceeds the reference value for the growth rate while
the amplitude stays below the reference value. In comparing the absolute values
from Phurbas to the other codes one shall to remember that the
Phurbas simulation is in three dimensions with a unstructured particle distribution.
However, at low resolutions the results for the growth rates are definitely
lower than that obtained in the grid codes. As the resolution is increased a
definite convergence towards the reference result is  observed.

From the mode amplitude plotted for \NDSPHMHD{} in, and notwithstanding the
aforementioned limitations to making comparisons in two dimensions, it is clear
that the cubic-kernel SPH is the least accurate method for the problem studied
in this work.  The result given here is however for a single initial
arrangement of SPH particles.  Results with SPH do depend, and in this problem
depend strongly, on the initial particle arrangement.  
In general, the \NDSPHMHD{} simulations show value and growth rates
for the $y$-velocity unstable mode amplitude which are too small.
At the lowest resolution ($128^2$) the simulation with artificial thermal conductivity
(Ne) gives a slightly improved result over the simulation without that addition (Nc),
but the dependence is minimal and more so at the higher resolutions.

The cubic-kernel SPH
results do not depend strongly on the use of a thermal conductivity term unlike
the sharp transition KHI test specified by \cite{2007MNRAS.380..963A}.  This is
demonstrated by the Nc simulation  where the thermal conductivity was turned off,
yielding results very similar to the Ne simulation.  This illustrates that the
artificial conductivity is not so much a patch for correcting Kelvin-Helmholtz
in SPH, but for ensuring that contact discontinuities stay well resolved.
Quintic kernel SPH, labeled as simulation No, uses  a larger number of neighbors and
has smaller zeroth-order SPH inconsistencies.  This gives a more accurate
result than cubic-kernel SPH for the same number of particles.  The pair of
\NDSPHMHD{} results Nc and No demonstrate that SPH converges in a limit that is
a combination of increasing particle number and neighbor number.  
The importance of using the quintic kernel over simply increasing the number of
SPH neighbor particles is to avoid particle clumping, which would effectively
lower the resolution, undermining the intent of a convergence study
\citep{2012JCoPh.231..759P}.  
Unlike in \cite{2010ARA&A..48..391S} we do not see
SPH starting at a acceptable growth rate at low resolutions and converging to a
lower growth late at high resolution.  We observe a much less surprising
behavior wherein the growth rate at low resolutions is too low, and the
solution appears to improve with increasing resolution, though the absolute
error is significant.

\subsection{Maximum $y$-Direction Specific Kinetic Energy}\label{sec_ydirener}

\begin{figure}
\begin{center}
\includegraphics[width=8cm]{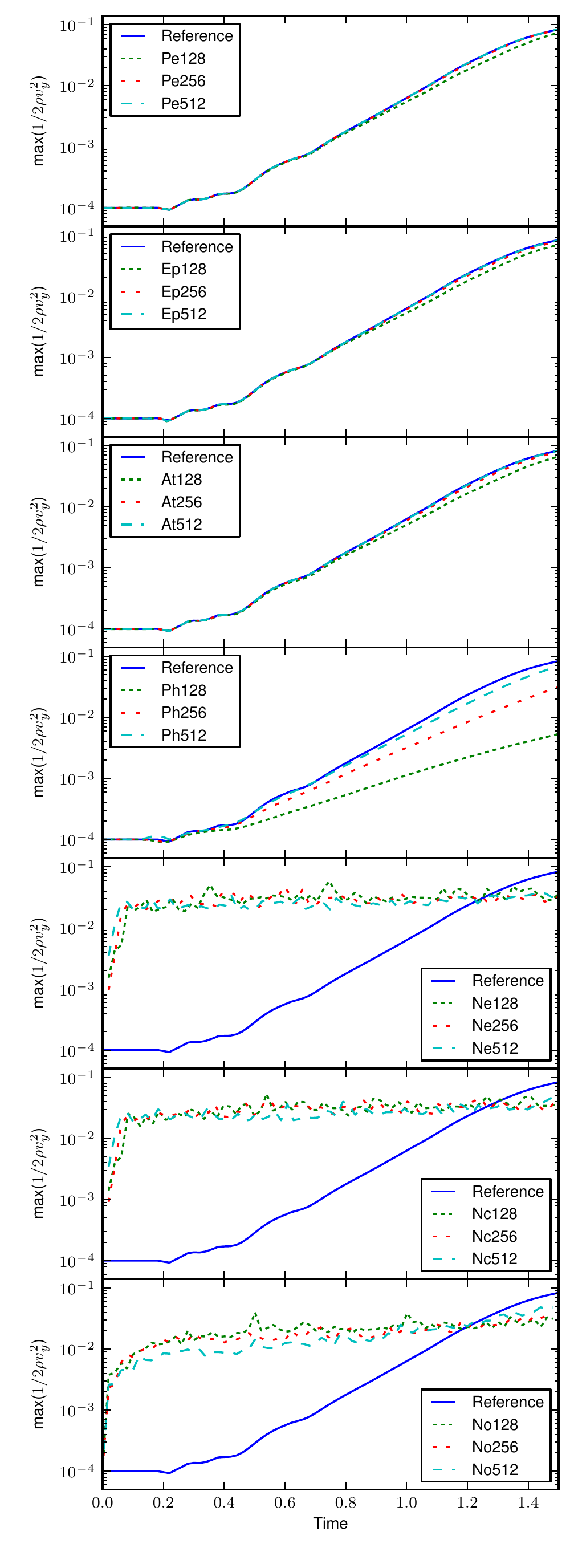}
\caption{Maximum $y$-direction kinetic energy in all codes.}
\label{allenergyplot}
\end{center}
\end{figure}

The behavior of the maximum $y$-direction kinetic energy is qualitatively
different from the mode amplitude as this measurement tracks the maximal value,
not a smooth average.  Maximum $y$-direction specific kinetic energy histories
are shown for all simulations in Figure~\ref{allenergyplot}.  Here the velocity noise
in SPH resulting from pressure force errors can be seen clearly in the overview
figure, while all other codes behave in a roughly similar manner.  The
convergence study does not establish an uncertainty on the maximum
$y$-direction kinetic energy, but the highest resolution Pencil Code result
plotted as the reference curve can be taken as a useful indicator of the
correct nonlinear solution.  

In Pencil, Enzo, Athena, and Phurbas, at late times at low resolution, when the
unstable velocity mode value is low, the maximum $y$-direction kinetic energy
is also low.  This is the opposite of the situation found in \NDSPHMHD{}, where
at late times at low resolution the unstable velocity mode value is low, but
the maximum $y$-direction kinetic energy is too high.

At lower resolutions in Phurbas the influence of velocity noise at the
interface can be clearly seen.  At early times the maximum $y$-direction
kinetic energy is too high and the unstable mode amplitude is too low.  Pencil
does not suffer from this to the same extent.  Enzo and Athena have the best
initial behavior at the interface as they are finite-volume schemes and hence
the initial pressure equilibrium is well represented across the interface. The
initial maximum kinetic energy and the initial mode amplitude are both to low
at low resolution in these codes.

The resolution dependence of the velocity noise is illustrated for the
cubic-kernel SPH with artificial conductivity (Ne).  Neglecting the artificial
conductivity yields virtually the same result as shown in
Figure~\ref{allenergyplot} (simulation Nc).  Quintic kernel SPH, with smaller zeroth-order
inconsistency errors than cubic-kernel SPH, does show smaller velocity noise,
but it is still very large (simulation No).

\subsection{Density at $t=1.5$}

\begin{figure*}
\begin{center}
\includegraphics[height=7in]{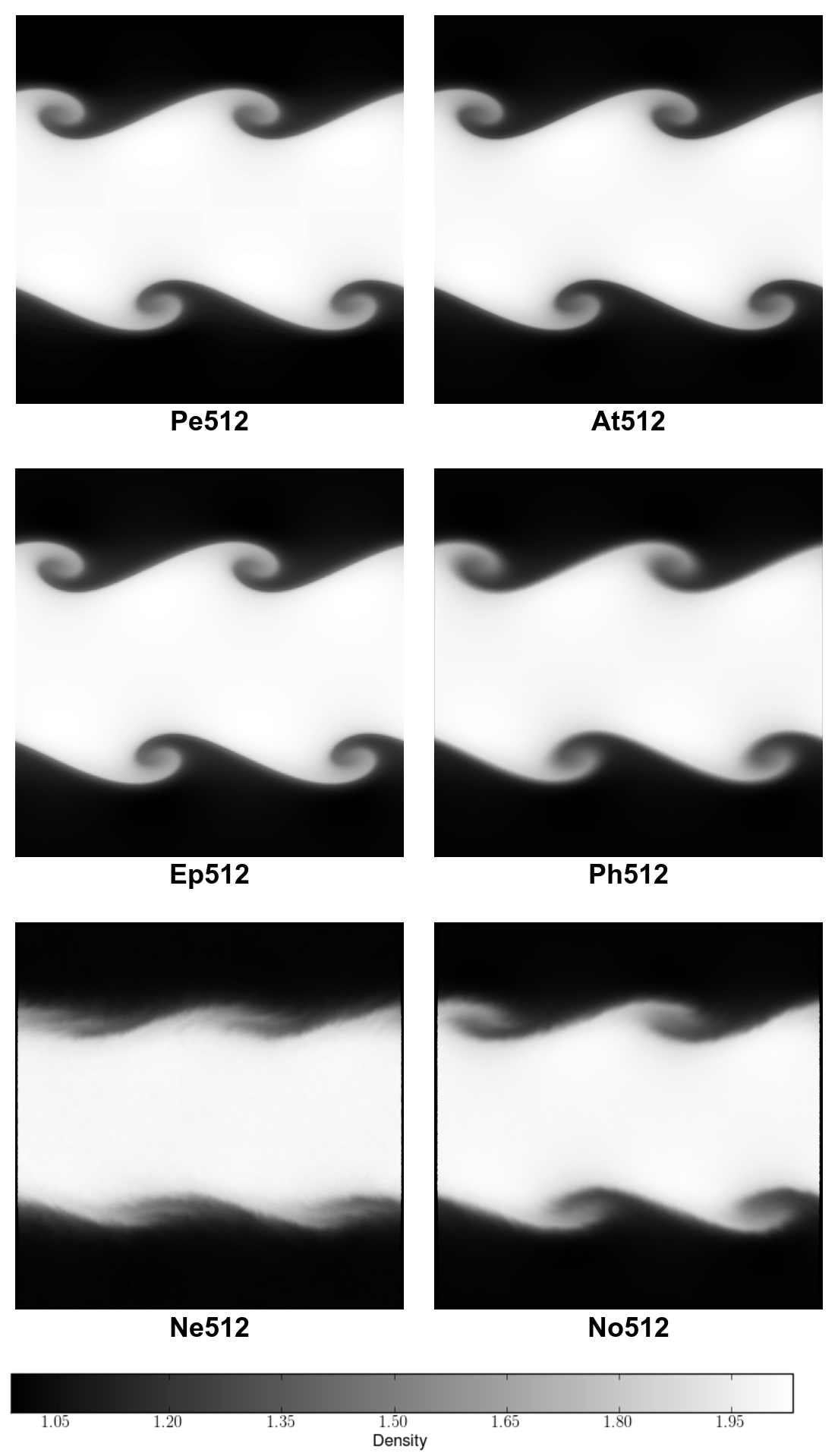}
\caption{Density at resolution $512^2$ and time $t=1.5$. {\em Upper Row:}
Pencil, Athena  {\em Middle Row:} Enzo, Phurbas {\em Lower Row:} \NDSPHMHD{}
cubic kernel, \NDSPHMHD{} quintic kernel} 
\label{figendrho512}
\end{center}
\end{figure*}

We show gray scale slices of the density field at $t=1.5$ in
Figure~\ref{figendrho512}.  All the images have the same limits on the grey
scale between density of $0.9883$ and $2.0320$, the density extremes in the
highest resolution result in the Pencil Code convergence study.  The results
for Pencil, Enzo, and Athena are largely similar, as at high resolution these
codes agree well with each other and with the reference result.  
Though the result with Phurbas strongly resembles the reference result 
although it clearly shows more diffusion.
The SPH
results from \NDSPHMHD{} (only Ne and No shown)  reflect the slow growth of the unstable $y$-velocity
mode already discussed.
The simulation No result using quintic-kernel SPH  shows less diffusion than simulation Ne using cubic-kernel SPH.
Especially in simulation Ne, secondary features of a filamentary appearance can be seen along the interface,
and these are less apparent in the simulation No result.
 The quintic kernel result (No) overall shows better agreement with
the reference than the cubic kernel result (Ne).

\section{Discussion} \label{sec_discussion}

Overall, the grid based codes Pencil Code, Athena, and Enzo had very similar
performance.  For these codes, the test problem in this work (run to $t=1.5$)
confirms their correctness.  This shows that the test as outlined here can be
used to discriminate among numerical schemes.  In this test, we demonstrated
that Phurbas and \NDSPHMHD{}, while both using meshless Lagrangian schemes,
give significantly different convergence behaviors.  Though Phurbas was run in
three dimensions, and \NDSPHMHD{} in two, the strikingly different qualitative
behavior bears some explanation.  A primary observation is that Phurbas differs
from \NDSPHMHD{} in that Phurbas uses a third order accurate and consistent
spatial discretization, while \NDSPHMHD{} uses an SPH discretization which has
zeroth-order inconsistency.  This issue is sufficiently complex that it is
discussed in a separate section (Section~\ref{sec_sph}).  We also note that no code
developed obvious signs of secondary instabilities in the solution by time
$t=1.5$, in agreement with the findings of the convergence study performed on
the reference result.  How, and when, secondary instabilities may arise in a KHI
test such as this is discussed in Section~\ref{sec_secondary}.

\section{The Behavior of SPH}\label{sec_sph}

\begin{figure}
\begin{center}
\plotone{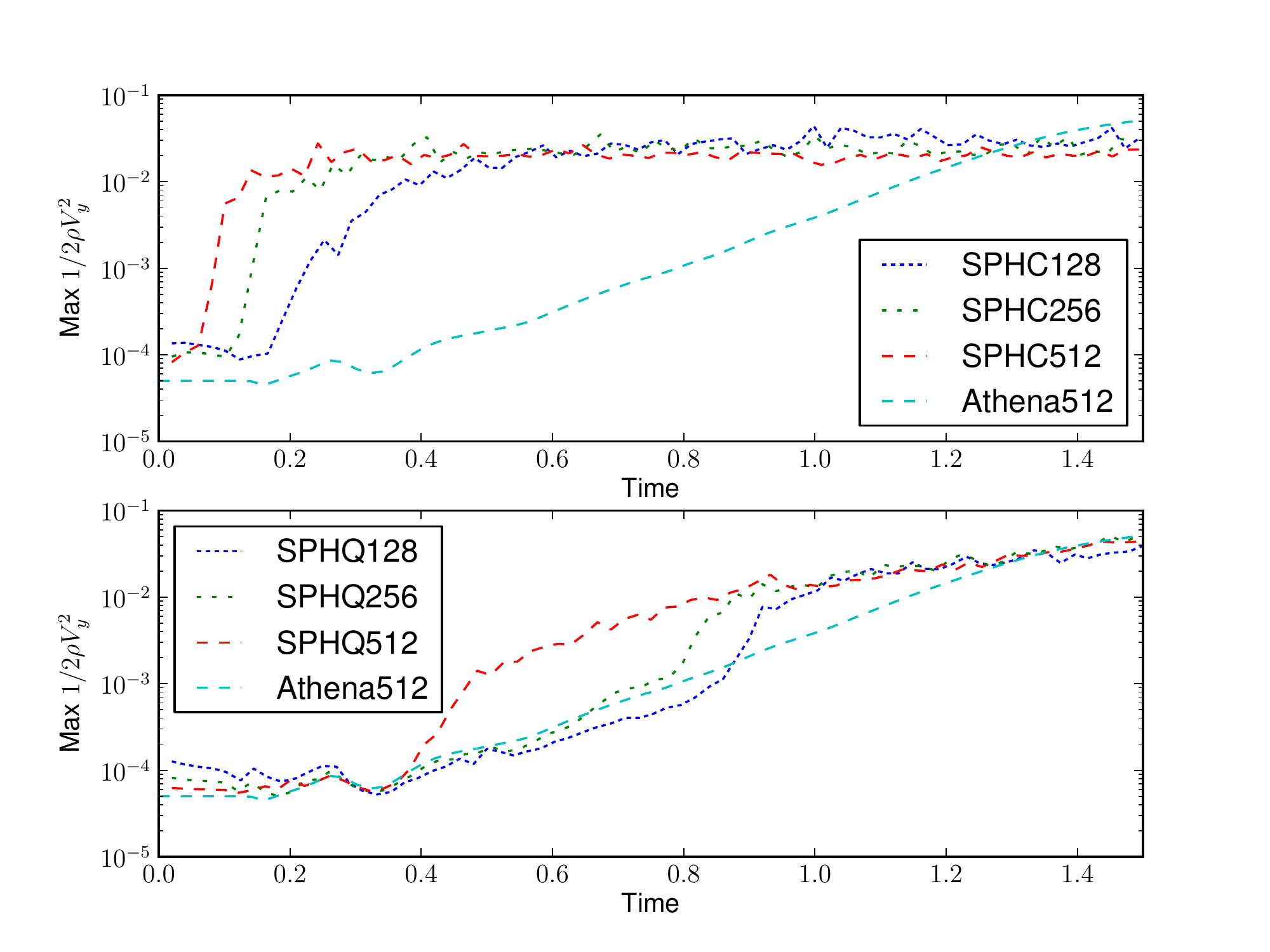}
\caption{$y$-direction kinetic energy in iso-density SPH, with a Athena result for comparison.
{\em Upper:} Cubic kernel SPH with \NDSPHMHD{}
{\em Lower:} Quintic kernel SPH with \NDSPHMHD{}
}
\label{sphisorho}
\end{center}
\end{figure}

\begin{figure}
\begin{center}
\plotone{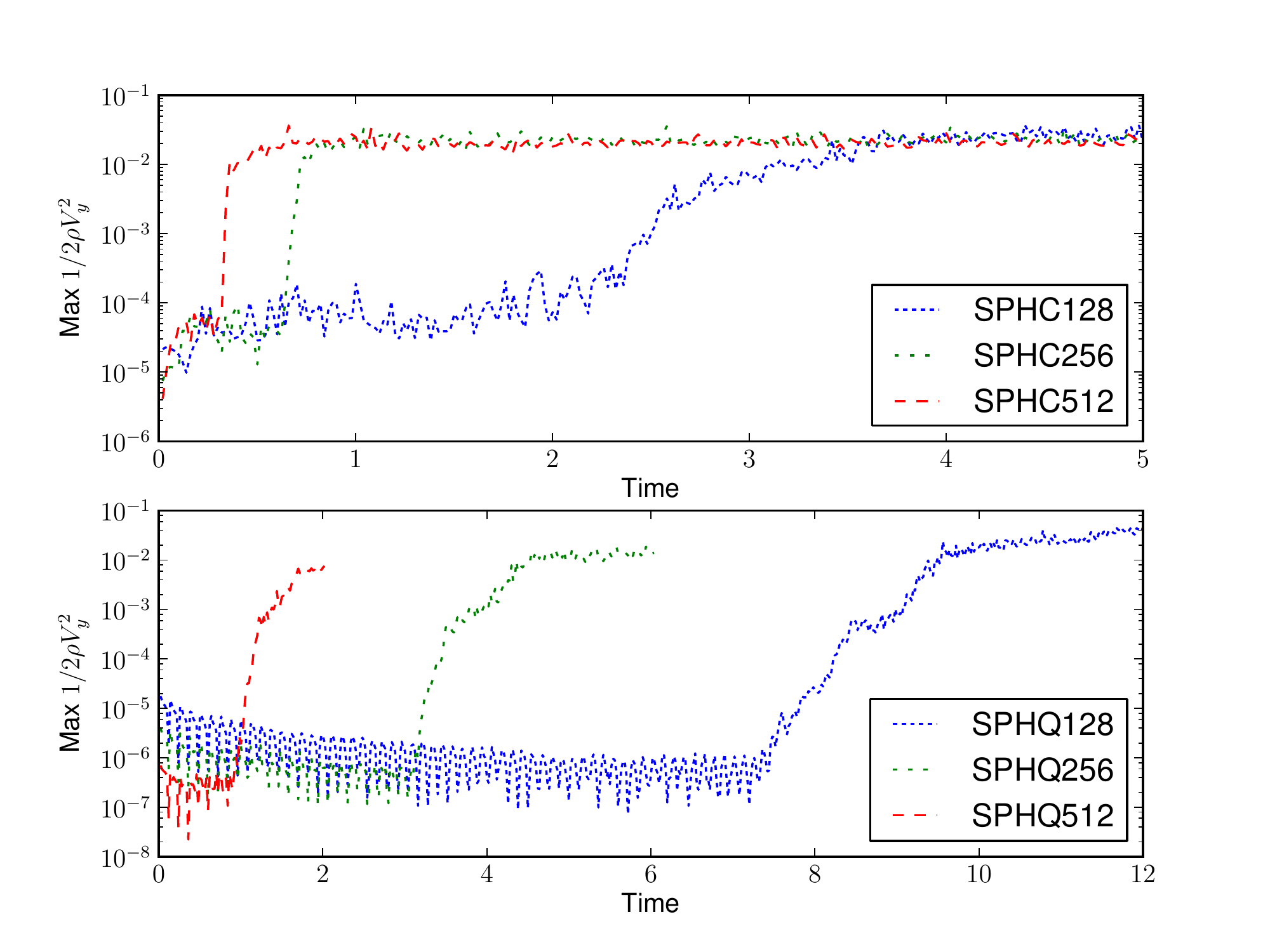}
\caption{$y$-direction kinetic energy in pure shear flow. 
{\em Upper:} Cubic kernel SPH with \NDSPHMHD{}
{\em Lower:} Quintic kernel SPH with \NDSPHMHD{}}
\label{sphshear}
\end{center}
\end{figure}

The results for the maximal $y$-direction kinetic energy in
Section~\ref{sec_ydirener} show significant noise appearing in the velocity in the
SPH simulations Ne, Nc, and No.  This section is devoted to exploring the source and
behavior of this noise.  It has been argued that maintaining particle order is
vital to achieving good results with SPH \citep{2011arXiv1111.1259P}.  Particle
ordering in SPH can be expressed as a condition that the Lagrangian of the
system of particles is minimized \citep[][Section 2.5]{2011arXiv1111.1259P}.
To seek this minimum, the particles must have some re-meshing motion in addition
to the pure fluid motions \citep[][Section 5.2]{2012JCoPh.231..759P}.  These
re-meshing motions mean that in SPH one always has some motions which are not
physical, but purely related to the SPH particles attempting to relax to an
ordered state \citep[][Section 5.2]{2012JCoPh.231..759P}.  These re-meshing
motions are also shown in the post-shock state in \citet[][Figure 10]{2012JCoPh.231..759P}.
  The re-meshing motions are provided by the linear errors in
the SPH pressure forces, which are in turn a result of the zeroth-order
inconsistency of SPH interpolation.  That is, the zeroth-order inconsistency in
SPH interpolation provides a linear error in the pressure force which causes
two particles which approach each other to repel.  Re-meshing motions created by
this repulsion are in turn damped by the artificial viscosity to encourage the
particle distribution to relax.  In this way, both the zeroth-order
inconsistency in the pressure estimate is vital to maintaining particle order
and the artificial viscosity cannot simply be disabled.  Further, though more
advanced artificial viscosities can be designed, the identification of the
particle velocity with the fluid velocity means that the need for motions
preserving particle order will necessarily corrupt to some degree the fluid
velocity itself.  The root cause that creates this situation is the zeroth-order
inconsistency in SPH interpolation, so the parameter which we vary is the one
which varies this error, the choice of SPH smoothing kernel.  As the change
from the cubic kernel to the quintic kernel decreases the size of the zeroth
order inconsistency, the re-meshing pressure forces are smaller, and the
resulting velocities are smaller.  Consequently, the level of $y$-direction
kinetic energy noise seen in the simulation No is smaller than that seen in the simulation 
Ne.

Recently \cite{2010MNRAS.403.1165C} and  \cite{2010MNRAS.405.1513R}
have connected zeroth-order inconsistency in SPH to
poor results for related KHI test problems, and  \cite{2011arXiv1109.4413B} has
demonstrated the connection in the context of low mach number turbulence. 
In this work we have demonstrated this
connection by first showing that on a smooth test problem the artificial
conduction of \cite{2008JCoPh.22710040P} does not significantly affect the
results.  Then, we have demonstrated that when the quintic kernel is used, with
smaller inconsistency errors than the cubic kernel, the test results improve
significantly.  The size of the inconsistency errors are reflected in the
maximal $y$-direction specific kinetic energy statistic, as pressure gradient
errors dive spurious particle motion.  Finally, Phurbas, while being meshless
and Lagrangian like SPH, uses a third-order consistent interpolation. The
consequence of this for the test in this paper is that it performs much better
than SPH, as the pressure forces are accurate enough to keep the velocity and
density noise much smaller than in SPH.  Hence, Phurbas has a qualitatively
different behavior on this test, and convergence does not depend on varying the
number of neighboring particles used in the interpolation.

To demonstrate that the velocity noise behavior seen in this work is general, we have
run a series of additional tests.  The first test is a version of the KHI setup
in Section~\ref{sec_setup} with a uniform density of 1.0.  For SPH, this is a
particularly simple choice as a uniform hexagonal close packed grid of
particles is the unique relaxed distribution in two dimensions.  Hence,
initially the setup does not suffer from any velocity noise.
Figure~\ref{sphisorho} shows that regardless of this initially relaxed
distribution, the maximal $y$-direction kinetic energy still reflects the
growth of the velocity noise.  Again, as in the previous tests the noise grows
sooner at higher resolutions.

We note that the velocity noise in the highest resolution quintic kernel case
of our isodensity test appears to be triggered by the growth of the primary KHI
instability.  To simplify the setup further, we remove the $y$-direction
velocity perturbation from the initial condition, yielding a smooth unperturbed
shear flow.  For the maximum $y$-direction specific kinetic energy measurement,
the trivial analytic solution for this problem is a value of zero for all
times.  Figure~\ref{sphshear}, upper panel, shows that in the this setup, run
with the cubic kernel, the maximal $y$-direction kinetic energy grows to the
same level as before in the isodensity KHI test, although it takes longer.
This growth happens at earlier times for higher resolutions.
Figure~\ref{sphshear}, lower panel, displays the same behavior for the quintic
kernel, although the timescales involved are longer.  
These simulations are stopped abruptly when they succumb to particle pairing instability 
\citep[][Section 5.4]{2012JCoPh.231..759P} and two particles approach within $10^{-8}$ length units.
Some recent proposed modifications to SPH  reduce or eliminate this instability for large kernels
\citep{2010MNRAS.405.1513R,2011arXiv1111.6985R,2012arXiv1204.2471D}.
With each kernel choice,
the time interval until the velocity noise jumps is shorter as the number of
particles is increased, but for a given number of particles the time interval
until the velocity noise jumps is longer if the quintic kernel is used. That
is, the results converge towards to analytic solution as the zeroth-order
inconsistency in the SPH interpolation is reduced.

\section{Secondary Instabilities} \label{sec_secondary}

\begin{figure*}
\begin{center}
\includegraphics[height=8in]{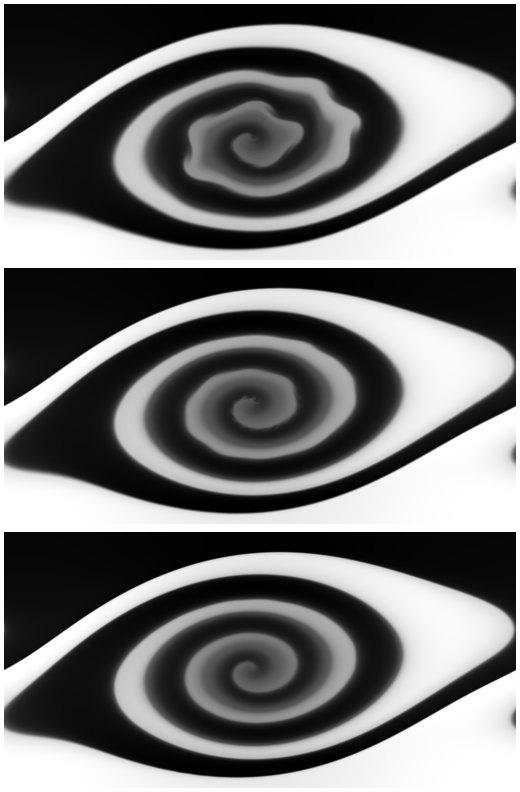}
\caption{Density in Athena at time $t=3.0$ at three resolutions, zoom-in on one primary KHI billow. 
Greyscale same as Figure~\ref{figendrho512}.
{\em Top:} $1024^2$
{\em Middle:} $2048^2$
{\em Bottom:} $4096^2$}
\label{figAtt30}
\end{center}
\end{figure*}

\begin{figure*}
\begin{center}
\includegraphics[height=8in]{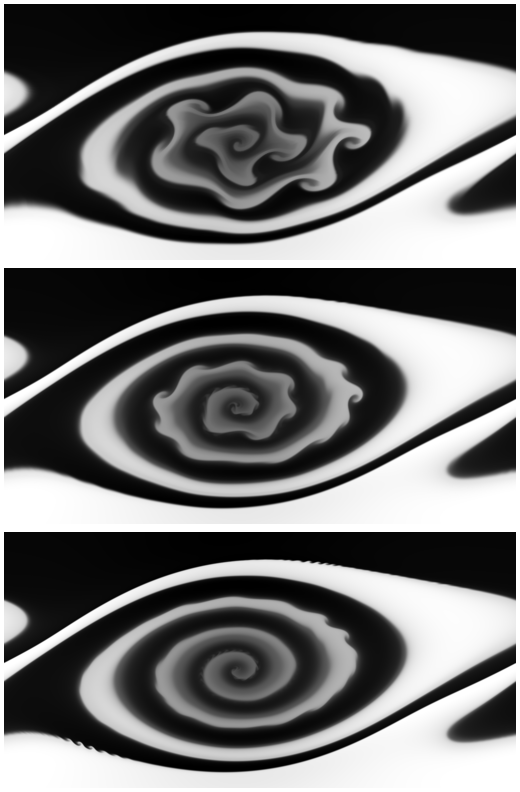}
\caption{Density in Athena at time $t=3.2$ at three resolutions, zoom-in on one primary KHI billow.
Greyscale same as Figure~\ref{figendrho512}.
{\em Top:} $1024^2$
{\em Middle:} $2048^2$
{\em Bottom:} $4096^2$}
\label{figAtt32}
\end{center}
\end{figure*}

\begin{figure}
\begin{center}
\plotone{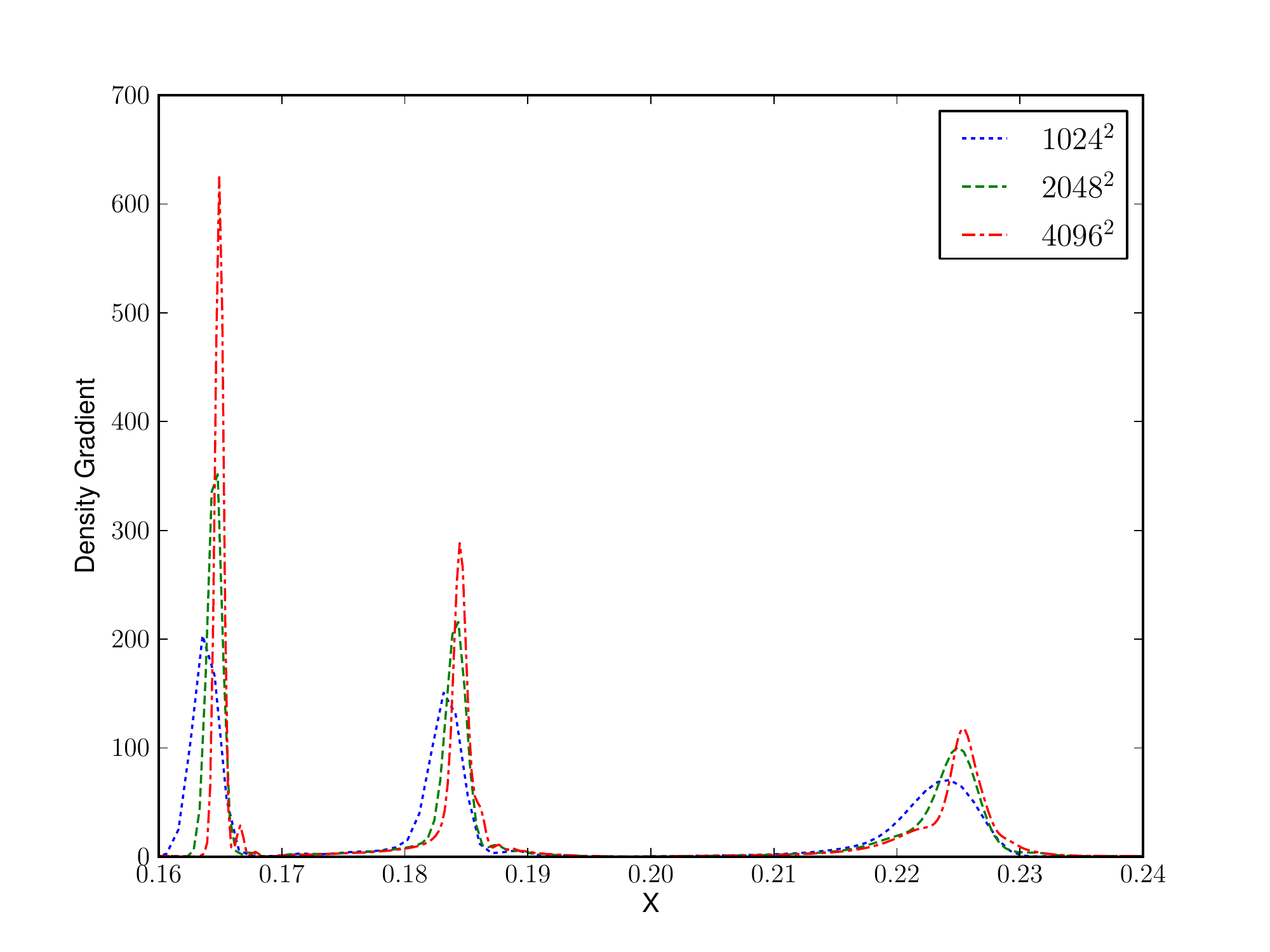}
\caption{Magnitude of the density gradient in Athena at time $t=3.0$ at three resolutions (denoted int the legend) at $y=0.125$ and
$x$ position as denoted on axis.}
\label{figAtrhograd30}
\end{center}
\end{figure}

We have shown that given a convergence test stated in a well posed manner, all
the methods tested appear to converge towards the correct result for the growth
of the primary instability.  Recent discussion of Kelvin-Helmholtz tests has
broadened to include secondary instabilities.  \cite{2011arXiv1109.2218S} shows
secondary instabilities developing from a similar initial condition.  The
reference solution we compute shows no indication of these structures.
\cite{2011arXiv1109.2218S} suggested that their moving-mesh code is able to
resolve secondary KHI billows which cannot be resolved in their fixed-mesh code
because it was too diffusive.  Though the solution in a fixed mesh and moving
mesh code should not be expected to by equivalent at any finite resolution,
that a given code does not develop secondary Kelvin-Helmholtz instabilities is
not simply a function of the diffusivity of the code, it is also a dependent on
the seeding of such instabilities. 

In general, we can categorize the possibilities for why our reference case and the 
tests done at lower resolutions in this work  do not show
the development of any secondary instabilities
in three cases:
\begin{enumerate}
\item The secondary billows should grow physically, either due to the nature of
the initial perturbation fed into the problem or the interaction between some
combination of the initial perturbation and modes of the instability directly
seeded by the initial perturbation. In this case the secondary billows should
eventually show up at some resolution in any convergent code, but should arise
at a particular location and time.

\item The secondary billows grow due to the balance of numerical perturbations
and numerical diffusion.  In this case the billows seen at some time should
disappear at some resolution in any convergent scheme as the significant power
in the numerical perturbations should eventually move to spatial scales too
small to seed the secondary instability efficiently.

\item The slight differences between the setup of \cite{2011arXiv1109.2218S}
and our setup mean the difference between seeing physical growth of secondary
billows and failing to produce them.  
\end{enumerate}
In the first case, developing the secondary instabilities is merely a matter of
using a sufficiently large resolution. At lower resolutions a
resolution study should still suggest a significant uncertainty or change
between simulations of different resolution as the secondary billows are damped less
and less.  In the second case, the resolution necessary to make the secondary
billows disappear may be quite large, as a numerical mechanism introducing noise
at a small scale  may still have significant power at larger wavelengths
depending on the spatial correlation of the mechanism introducing the noise.
However, the resolution study
performed with the Pencil Code to much higher resolution shows no indication that these
modes grow.  
The third case cannot be ruled out explicitly, as \cite{2011arXiv1109.2218S}
does not specify the exact details of the setup used. However, we can show that
for our problem that secondary instabilities that do develop are of a purely
numerical origin.  This strongly suggests that the secondary billows seen in
\cite{2011arXiv1109.2218S} are a numerical artifact, so the observation that a
fixed grid codes does not develop these on the same problem does not imply that
the fixed grid code is too diffusive to support the modes.

To demonstrate how numerical effects can seed secondary KHI we have
performed a test with Athena.  We ran the KHI test at resolutions of $1024^2$,
$2048^2$ and $4096^2$ until $t=4$.  In Figure~\ref{figAtt30} the density in a
region centered on a single primary KHI billow is shown at time $t=3$.
Secondary KHI billows can be seen growing in the $1024^2$ case, and this pattern
is successively suppressed at the higher resolutions, suggesting it is an
artifact of the finite resolution and is converging away.  However, in the
$2048^2$ resolution, a different set of secondary instabilities can be seen
growing at much shorter wavelengths in the central winding of the primary
billow.  As the resolution is increased, the numerical seeding of the secondary
instabilities changes, and the secondary modes which are excited change.
Figure~\ref{figAtt32} shows the same region at time $t=3.2$.  By this point,
the secondary instabilities in the $4096^2$ resolution simulation have become
apparent.  Surprisingly, a new set of secondary billows has appeared on the
outer winding of the primary billow.  We cannot reach well justified
conclusions about a particular mode of the secondary instability from this
study as we cannot reproduce the same instability at two different resolutions.

Though the growth of secondary instabilities is likely  a physical reality at
the Reynolds numbers involved in astrophysical problems, relying on numerical
effects to seed them will not result in a true physical model of the phenomena
as the seeding, and hence growth of these instabilities, will be inherently
dependent on parameters such as resolution.  Models relying on numerically
seeded instabilities, even if the presence of the instability is physical, make
it difficult to seperate numerical effects from physical behavior, which is
turn makes it difficult to come to strong conclusions about the effect of the
instability. 
 Conclusions about the instability
must consist of a measurement and some manner of characterization of the error
in that measurement.  In order to characterize the error in the model of the
instability, a convergence study must be performed.  To perform this
convergence study, a fixed seeding of the instability must be possible across
all resolutions.  If the seeding is a numerical effect caused by the finite
resolution, it will not be fixed between two  resolutions.  Hence, the tests in
this work do not show that any code used cannot, at any of the the resolutions
tested, resolve secondary Kelvin-Helmholtz instabilities as these have not been
seeded in a controlled way or even in an avoidable manner.

From the results of the convergence study in Section~\ref{sec_reference} we propose
that if a code develops secondary Kelvin-Helmholtz billows in this test by
$t=1.5$, it is due to the growth of numerical perturbations.  The less rigorous
study performed with Athena suggests that the same conclusion should hold to at
least $t=2.5$.  In the limit of infinite resolution, any convergent code should
reproduce the correct result.  However, if at finite resolution a code shows a
tendency to produce secondary instabilities, then the scheme can be improved by
adding a diffusive operator to damp the noise leading to the the instability.
Particularly with respect to moving-mesh tessellation codes, Kelvin-Helmholtz
tests are not the only place where behavior suggests that some additional
numerical diffusion should be used to damp grid scale noise.

Development of secondary Kelvin-Helmholtz instability after $t=1.5$ is likely
due to the presence of spurious noise in the solution.  For example, in the
preparation of this work, we discovered that the evolution of the test problem
here differed greatly at high resolution ($4096^2$) between Enzo versions 1.5
and 2.0.  In Enzo 2.0, a bug existed that caused slightly incorrect pressure
reconstructions.  This caused small sound waves to launch from the interface,
and propagate though the periodic domain interacting with themselves and
forming small short-wavelength perturbations. 
The discovery of this bug was fortuitous however, because it demonstrates
again how artificial, numerical perturbations can give rise to secondary
Kelvin-Helmholtz in this test problem if they are able to overwhelm the
dissipation of the scheme.

The underlying cause of the tendency of many schemes to develop secondary KHI in
this test problem is that the shear interface becomes increasingly steep as it
stretches in the primary KHI billow.  Eventually, the width of the interface
approaches the grid scale and it become susceptible to numerically seeded
secondary instabilities.  This behavior is also commonly seen in the initial
evolution of  grid-aligned sharp transition versions of the KHI test.  Two
examples are \citet[][Figure 13]{2010MNRAS.407.1933J} and
\citet[][ Figure 8, upper right panel]{2011arXiv1109.2218S}.  We suggest then
that even fixed-grid finite volume Godunov schemes may be improved in
pathological cases of unresolved shear interfaces by the addition of a
diffusive flux.  This flux should be chosen to spread the interface
over enough grid cells to suppress the numerically seeded instabilities.

Another lesson to be derived here is a cautionary one.  Not all new
instabilities seen as resolution is increased when solving the discretized
Euler equations are physically real. New numerical instabilities can reveal
themselves as resolution is increased, as the flow can enter into new regimes
where it is more sensitive to the inevitable numerical noise in a method.  In
the high resolution set of Athena simulations, this can be seen in the magnitude of
the density gradients.  In Figure \ref{figAtrhograd30} the density gradient in
a slice through a primary billow at time $t=3.0$ in the three Athena simulations used
in this section is plotted, calculated with a four point second-order finite
difference stencil.  As the resolution is increased, the maximum gradient
achieved increases.  Mathematically, when solving the Euler equations, this
behavior arises because the modified equations that are actually solved by the
method change as resolution is increased - the diffusive effects become
smaller.  One route around this difficulty can be to solve the Navier-Stokes or
Boltzmann equations instead with a fixed viscosity or particle mean free path.
Since these equations have a physical scale where diffusion dominates dynamics,
the reliable elimination of numerically generated instabilities for arbitrarily
long run times can be obtained by fixing the physical diffusive scale and
reducing the grid scale far below the diffusive scale.

The same point illustrates how the transition to turbulence and mixing must be
studied when the Euler equations are used.  To produce the secondary
instabilities that break up the flow, the nonlinear interaction or modes or
seeding of secondary instabilities must be done on a controlled manner.  This
job cannot be left to the numerical noise, or the time and manner in that the
flow breaks up will  be a reflection of numerical issues and not of physical
reality.

Finally, we suggest that it is possible to produce a controlled test of the
growth of secondary billows from definite perturbations, similar to the study
performed by \cite{2008JFM...612..237F} in an incompressible flow.  Such a
setup could be useful in determining the appropriate and minimal diffusion to
add to a scheme to suppress the numerical seeding of secondary instabilities in
given conditions.

\section{Conclusions}\label{sec_conclusions}

We have constructed a reference solution with a well characterized uncertainty,
along with defining a general manner in that the test can be analyzed.  This
methodology was applied to example codes from the major families of numerical
techniques used in astrophysics.  All codes tested showed convergence towards
the reference result when the resolution was increased in the appropriate
manner.  For SPH, the use of an artificial thermal
conductivity does not significantly effect the results, but using a
higher-order kernel (and hence a larger number of neighbors) does improve the
results.  We conclude then that the fundamental reason for poor performance of
SPH in KHI is the zeroth-order inconsistency of SPH interpolation.  Visually, to
time $t=1.5$ in the test problem there are no secondary instabilities that
arise in the reference solution.  By examining the relative behavior of
different types of code, we argue that the presence of secondary instability on
this test is caused by having a numerical diffusion that is very low compared
to the grid noise in the method.  Hence, we propose that it is advantageous in
some methods, particularly moving-mesh tessellation methods, but also in
fixed-grid Godunov schemes, to include an extra diffusion operator to smooth
the solution such that grid noise does not drive small scale instabilities.

\acknowledgments
We are indebted to the authors of \NDSPHMHD{}, Athena, Enzo and the Pencil Code
for making the codes freely available to us.  The SPH visualizations were
created with {\sc SPLASH} \citep{2007PASA...24..159P}.  We thank the anonymous
referee for constructive comments that significantly improved the organization
of the paper.  We thank Daniel Price for feedback on a draft of this
manuscript, and suggesting the use of fully relaxed SPH initial conditions and
the iso-density special case.  We thank
Mordecai-Mark~Mac~Low for his support and useful discussions.  We thank Paul
Duffel for useful discussions about moving-mesh schemes.
J-C.P.\ thanks Orsola De~Marco and Falk Herwig for their support.  This work
has been supported by National Science Foundation (NSF) grants AST-0835734 and
AST-0607111.  W.L.\ gratefully acknowledges partial financial support from the
NSF under grant no. AST-1009802. W.L.\ completed co-writing this work at the
Jet Propulsion Labratory, California Institute of Technology, under a contract
with the National Aeronautics and Space Administration.  This work used the
Extreme Science and Engineering Discovery Environment (XSEDE), which is
supported by National Science Foundation grant number OCI-1053575.  This work
used computer time provided partially by Westgrid and Compute Canada.

\end{document}